# Topological phases in pyrochlore thallium niobate $Tl_2Nb_2O_{6+x}$

Wei Zhang,[*,1,2] Kaifa Luo,[3] Zhendong Chen,[1,2] Ziming Zhu,[4] Rui Yu,[‡,3] Chen Fang,[5,6,7,8] Hongming Weng[†,5,6,7,8]

[1]*Fujian Provincial Key Laboratory of Quantum Manipulation and New Energy Materials, College of Physics and Energy, Fujian Normal University, Fuzhou 350117, China*

[2]*Fujian Provincial Collaborative Innovation Center for Optoelectronic Semiconductors and Efficient Devices, Xiamen, 361005, China*

[3]*School of Physics and Technology, Wuhan University, Wuhan 430072, China*

[4]*Key Laboratory of Low-Dimensional Quantum Structures and Quantum Control of Ministry of Education, Department of Physics and Synergetic Innovation Center for Quantum Effects and Applications, Hunan Normal University, Changsha 410081, China*

[5]*Beijing National Laboratory for Condensed Matter Physics, and Institute of Physics, Chinese Academy of Sciences, Beijing 100190, China*

[6]*University of Chinese Academy of Sciences, Beijing 100049, China*

[7]*Songshan Lake Materials Laboratory, Dongguan, Guangdong 523808, China*

[8]*CAS Centre for Excellence in Topological Quantum Computation, Beijing 100190, China*

Email: [*]zhangw721@163.com, [‡]yurui@whu.edu.cn, [†]hmweng@iphy.ac.cn

**ABSTRACT**

The discovery of new topological electronic materials brings a chance to uncover novel physics and plays a key role in observing and controlling various intriguing phenomena. Up to now, many materials have been theoretically proposed and experimentally proved to host different kinds of topological states. Unfortunately, there is little convincing experimental evidence for the existence of topological oxides. The reason is that oxidation of oxygen leads to ionic crystal in general and makes band inversion unlikely. In addition, the realization of different topological states in a single material is quite difficult, but strongly needed for exploring topological phase transitions. In this work, using first-principles calculations and symmetry analysis, we propose that the experimentally tunable continuous solid solution of oxygen in pyrochlore $Tl_2Nb_2O_{6+x}$ ($0 \leq x \leq 1.0$) leads to various topological states. Topological insulator, Dirac semimetal and triply degenerate nodal point semimetal can be realized in it via changing the oxygen content and/or tuning the crystalline symmetries. When $x = 1$, it is a semimetal with quadratic band touching point at Fermi level. It transits into a Dirac semimetal or a topological insulator depending on the in-plane strain. When $x = 0.5$, the inversion symmetry is spontaneously broken in $Tl_2Nb_2O_{6.5}$, leading to triply degenerate nodal points. When $x = 0$, $Tl_2Nb_2O_6$ becomes a trivial insulator with a narrow band gap. These topological phase transitions driven by solid solution of oxygen are unique and physically plausible due to the variation of valence state of $Tl^{1+}$ and $Tl^{3+}$. This topological oxide will be promising for studying correlation induced topological states and potential applications.

**INTRODUCTION**

The remarkable discoveries of various quasiparticles in solids with or without the counterpart in high-energy physics have inspired intensive studies on topological electronic materials (TEMs).[1-5] They are promising for future applications, owing to low-dissipation transport property and intrinsic insensitivity to environment perturbations. TEMs are characterized as having electronic structures with non-trivial topology in momentum space. Typically, TEMs can be classified into topological insulator (TI),[6-8] topological semimetal (TSM)[9-14] and topological superconductor. The initial impetus originates from the TI, which exhibits linear dispersive surface/edge states and can make novel quantum electronic devices compatible with current electronic technologies. Moreover, magnetically doped TIs are proved to hold quantum anomalous Hall effect.[15,16] Recently, research focus of TEMs has shifted towards TSMs, which have exotic transport properties.[17-19] TSMs are special metals with Fermi surfaces composed of and only of nodal points. They include four members, namely Weyl semimetal (WSM),[20,21] Dirac semimetal (DSM),[10,11] nodal line semimetal (NLSM)[22,23] and triply- or multiply- degenerate nodal point (TDNP or MDNP) semimetal.[24-29] These TSMs are distinguished from each other by the degeneracy of the nodal points and the topological protection mechanism. WSM has isolated double-degenerate nodal points at or close to the Fermi level and is topologically robust as long as the translation symmetry of lattice is preserved, while DSM has isolated four-fold degenerate nodal points and is protected by proper crystalline symmetries. NLSM contains continuous nodal points forming lines, while MDNPs host three-, six- or eight-fold degenerate nodal points. Both of them need some proper crystalline symmetries, such as rotation, mirror and/or nonsymmorphic translation. The TDNP in WC family is a crossing point formed by a nondegenerate band and a double-degenerate band.[25,27] It is identified as an intermediate state between Weyl and Dirac TSM, bringing in new interesting physics. However, there have been quite few reports on TEMs discovered in oxide materials till now[12,30-34] and their properties are to be extensively explored once they are available experimentally.

$Tl_2Nb_2O_{6+x}$ is in pyrochlore structure,[35] which has been known since 1960s. The ideal pyrochlore $Tl_2Nb_2O_6O'_{x=1}$ was first discovered, and then Fourquet *et al.* demonstrated that there exist continuous solid solutions $Tl_2Nb_2O_6O'_x$ ($0 \leq x \leq 1.0$) via thermogravimetric analysis (TGA), chemical analysis, and X-ray thermodiffractometry.[36] Interestingly, with the removal of (1-x) O' out of the $Tl_2Nb_2O_7$, the Tl atoms could shift along [111] axis and be away from the central symmetric position, leading to spontaneous inversion symmetry breaking, which brings a very unique way to systematically tune the topological phases in it.

In this work, we propose that $Tl_2Nb_2O_{6+x}$ can have several attractive topological features as x changes. DSM, TI and TDNP semimetal states all can be realized in $Tl_2Nb_2O_{6+x}$ series via tuning the crystalline symmetry or oxidation level. When x = 1, $Tl_2Nb_2O_7$ is cubic and is a zero-gap semimetal similar to HgTe with quadratic contact point (QCP) at Γ.[37] In-plane compressive and tensile strain can drive it into DSM and TI, respectively. When x = 0.5, $Tl_2Nb_2O_{6.5}$ has no inversion symmetry and is a TDNP semimetal. When x = 0, $Tl_2Nb_2O_6$ is a trivial insulator with narrow band gap. Because strain engineering greatly contributes to exploring physics[37-40] and quite a small strain is introduced here, it is feasible for experimental observation of the topological states in $Tl_2Nb_2O_{6+x}$. Moreover, many intriguing phenomena and rich physics have been found in pyrochlore oxides, such as complex magnetic phases, superconducting and multiferroics. Thus, our studies may provide a unique platform for investigating the strongly correlated topological phases, multi-phase control and potential applications.

## RESULTS AND DISCUSSION

## Crystal structure

The ternary oxide $Tl_2Nb_2O_{6+x}$ belongs to the pyrochlore structure. The ideal structure of x = 1 is in space group *Fd-3m* (No. 227) (Fig. 1a), whose first Brillouin zone (BZ) is shown in Fig. 1b. Tl and O' atoms are located at 16*d* (1/2, 1/2, 1/2) and 8*b* (3/8, 3/8, 3/8) positions, respectively. Four Tl atoms will form a tetrahedra with an O' atom at the center. Nb and O atoms are in 16*c* (0, 0, 0) and 48*f* (0.2925, 1/8, 1/8) positions, respectively, forming $NbO_6$ octrahedra. The experimental lattice constant is $a_0$ = 10.622 Å and is used for the calculations in the present paper.[36]

Compared with the case of x = 1, the missing of O' makes the same number of Tl ions change from +3 to +1 and shift away from the centrosymmetric position (Fig. 1c). Though the distribution of O' vacancy and $Tl^{+1}$ ions is somehow random in x = 0.5 case, we take away half of O' atoms in the primitive unit cell (*Z* = 2), and Tl atoms are shifted away from 16*d* to the 32*e* (0.507, 0.507, 0.507) positions. The lattice constant is taken as 10.6397 Å according to the experimental value in x = 0.490 case, which is the closest to 0.5.[36] The crystal structure symmetry becomes *R3m* (No. 160) without inversion center, being different from that of $Tl_2Nb_2O_7$. When x is reduced to 0, all the Tl atoms becomes +1 and stay on the noncentral position. The lattice constant is taken as 10.6829 Å, which is the experimental value when x = 0.070. [36]

## Band structure of bulk $Tl_2Nb_2O_7$

The 5*d* orbitals of $Tl^{3+}$ atom split into $e_g$ and $t_{2g}$ orbitals due to the crystal field formed by oxygen hexagonal bipyramid. Without considering the SOC, $Tl_2Nb_2O_7$ is a QCP semimetal with a triply degeneracy at Γ point (Fig. 1d), which is also verified by the hybrid functional HSE06 calculation (red color bands in Fig. 1d). This is the same as the results in Materiae, an online database of topological materials, and other similar databases.[41-43] The states at Γ point mainly come from $t_{2g}$ orbitals composed by hybridization of Tl 5*d* and O 2*p* orbitals. When SOC is taken into consideration, SOC splitting among *p* orbitals is opposite to that among $t_{2g}$ orbitals.[44] Therefore, the final effective SOC of the Γ point is determined by the competition between Tl $t_{2g}$ and O *p* spin-orbit splitting.[44] With SOC, the QCP at Γ (Fig. 1d) splits into a double degenerate $\Gamma_7^+$ band and a fourfold degenerate $\Gamma_8^+$ states. (Fig. 1e, 2c) $\Gamma_7^+$ is higher than $\Gamma_8^+$, which indicates that the effective SOC in these bands is negative due to the *d-p* hybridization as discussed in TlN.[44] The fourfold degenerate $\Gamma_8^+$ is half occupied and becomes another QCP similar to HgTe.[37]

**Band structure and topological property of strained $Tl_2Nb_2O_7$**

The QCP at Γ is protected by $O_h$ point group. Breaking the $O_h$, this fourfold degeneracy will be lifted, and thus topological insulating states or topological semimetal states are formed.[37] In this section, we consider the topological phase transition in $Tl_2Nb_2O_7$ system with strain (positive strain refers to expansion, while negative strain refers to compression). The related space group is changed from *Fd-3m* (No. 227) to *I4₁/amd* (No. 141). A top view of the structure without strain is shown in Fig. 2a, while its non-SOC and SOC bands are shown in Fig. 2b, 2c for comparison. From the pictures, we can see when SOC is included, the gapless semimetal is formed owning to the fourfold degeneracy of $\Gamma_8^+$, which is also similar to the case of $Cu_2Se$.[45] In $Tl_2Nb_2O_7$, $\Gamma_7^+$ states are higher than $\Gamma_8^+$ states, while in $Cu_2Se$, $\Gamma_8^+$ states are higher than $\Gamma_7^+$ states.

A compressive strain of -1% in xy-plane is applied (Fig. 2d) and the lattice constants become $a$ = b = 0.99 $a_0$, and $c$ = 1.02 $a_0$. The band structures without and with SOC are calculated and compared in Fig. 2e, 2f. When SOC is neglected, the application of strain breaks the system symmetry and results in the point group changing from $O_h$ to $D_{4h}$ at Γ point. This leads to orbitals like $d_{xz}$ and $d_{yz}$ extending in $z$ direction and having different on-site energy from that of in-plane orbital like $d_{xy}$. The valence-band maximum and conduction-band minimum are degenerate at Γ point, and the third band crosses them along Γ-Z direction. Wave-function analysis explains that these three bands are described by different irreducible representations (IRs) of the $C_{4v}$ point group. The IR of the band shown in black is $B_2$, while that of the double degenerate bands shown in red and blue is $E$.[46] Thus, in the non-SOC case without the spin degree of freedom, two TDNPs related with inversion or time-reversal symmetry can be formed in the -Z to Γ and Γ to Z directions, respectively. The energy band with SOC in strain of -1% case is also calculated (Fig. 2f). SOC drives a phase transition from the QCP semimetal to Dirac semimetal, where two fourfold degenerate Dirac points are on the path -Z to Γ and Γ to Z, respectively. Detailed wave-function analysis shows the bands forming Dirac cone in black/red and blue/green belong to different IRs of $C_{4v}$ point group: $E_{1/2}$ and $E_{3/2}$.[46] To further prove the topological properties of this phase, Wilson loop method[47] is leveraged to trace the evolution of the Wannier charge centers in $g_3 = 0$ and π planes. As shown in Fig. S1 (see Supplementary Information), there exists one crossing of Wannier center (black lines) and the reference line (red line) in $g_3 = 0$ plane, but not for $g_3 = \pi$ plane, confirming that $Tl_2Nb_2O_7$ with strain of -1% is a topological Dirac semimetal with $Z_2 = 1$. Moreover, surface states on (010) surface are calculated based on the tight-binding Hamiltonian constructed with Wannier functions using Green's-function method. Both bulk 3D Dirac point and gapless nontrivial surface states are clearly present, which makes the $Tl_2Nb_2O_7$ with -1% strain fantastic for exploring the coupling between Dirac point and topological insulator states (Fig. 3a). There are two branches of surface states emerging in the gap and touching at $\bar{X}$ and $\bar{Z}$ points due to Kramer's degeneracy. One branch connects to the conduction bulk bands, while the other one links the valence bulk bands. Moreover, iso-surface with energy at Dirac points of the surface states is calculated (Fig. 3b). There exists a pair of surface Fermi arcs connecting two projected Dirac nodes.

An expansion strain of 1% in the xy-plane is applied to the system (Fig. 2g), and the related lattice constants are changed to $a$ = b = 1.01 $a_0$, and $c$ = 0.98 $a_0$. The non-SOC band structure is shown in Fig. 2h. There exists one band intersection along the X-Γ, which is formed by bands in red and blue. The two bands belong to IRs of the $C_{2v}$ point group: $A_2$ and $B_2$, respectively.[46] In fact, this nodal point is on a nodal line in $k_x$-$k_y$ plane, which is protected by the coexistence of inversion and time-reversal symmetries (see Supplementary Information Fig. S2). When SOC is included, the gap is fully opened in the entire nodal line (Fig. 2i), generating a strong TI with global gap of ~13 meV. The same Wilson loop method is used here to identify the topological property of the structure with strain of 1% (see Supplementary Information Fig. S3). Topologically protected surface Dirac cone on (010) surface connecting the conduction and valence bands emerges inside the gap (Fig. 3c). These two branches of surface states also touch at $\bar{X}$ and $\bar{Z}$ points, similar to the case of -1% strain. Furthermore, iso-energy plot of surface states at energy of -6 meV in the gap is displayed (Fig. 3d).

To understand the phase transition mechanism under strain, an effective $k \cdot p$ model is constructed (see Supplementary Information). From Fig. S4, we can see the band structures coincide with those from first-principles calculations.

**TDNPs in noncentrosymmetric $Tl_2Nb_2O_{6.5}$**

Compared with the case of x = 0, the extra O' (in the network of $Nb_2O_6$) oxidizes one of the nearest four Tl atoms to +3 and repels the other three monovalent Tl. Therefore, Tl atoms are away from the centrosymmetric position. Such natural breaking of inversion symmetry provides a new material hosting intrinsic TDNPs, whose topological properties are studied by calculating the band structure with SOC (Fig 4). We can see band crossings along Γ-L ([111] axis) host massless fermions, which appear in a pair due to time-reversal symmetry. To shed light on the forming mechanism of TDNPs, wave-function analysis is performed. These bands belong to IRs of the $C_{3v}$ point group: the IRs of the black/green, blue and red bands are $E_{1/2}$, $2E_{3/2}$, $1E_{3/2}$, respectively.[46] Thus, band crossings can form two TDNPs near Γ point, protected by $C_{3v}$ and time-reversal symmetries. HSE06 calculation without SOC also confirms the existence of TDNP along Γ-L direction (see Supplementary Information Fig. S5). These TDNPs are about 0.4 eV below Fermi level, which is higher than those in MoP[28] while lower than those in WC.[29].

To understand the mechanism forming TDNPs in $Tl_2Nb_2O_{6.5}$, we further introduce a $k \cdot p$ model around Γ point with SOC included. In order to make the discussions simple, we chose the [-110] and [111] direction (Fig. 1b) as $k_a$ and $k_c$ axis, respectively. The little group at Γ point is $C_{3v}$, which includes two generators: a $C_3$ rotation about $k_c$ axis and a mirror symmetry $M_{100}$ with the normal of the mirror plane in $k_a$ direction. The Hamiltonian of $k \cdot p$ model keeping invariant under $C_3$ and $M_{100}$ symmetries is obtained as

$$H(\vec{k}) = \begin{pmatrix} e_{10} + e_{11}k_- k_+ + e_{12}k_c^2 & h_{121}k_- & h_{132}k_+ & h_{141}k_c \\ h_{121}k_+ & e_{20} + e_{21}k_- k_+ + e_{22}k_c^2 & h_{221}k_- & -h_{132}k_+ \\ -h_{132}k_- & h_{221}k_+ & e_{20} + e_{21}k_- k_+ + e_{22}k_c^2 & h_{121}k_- \\ h_{141}k_c & h_{132}k_- & h_{121}k_+ & e_{10} + e_{11}k_- k_+ + e_{12}k_c^2 \end{pmatrix}$$

up to the first order of $\boldsymbol{k}$ with the basis set in order of |+3/2>, |+1/2>, |-1/2>, |-3/2>, which are the four eigenstates at Γ point near the Fermi level $k_\pm = k_a \pm ik_b$. The coefficients $e_s$ and $h_s$ are parameters that can be obtained by fitting the first-principles results, listed as Table S1. The bands

comparison between the $k \cdot p$ model and the first-principles calculations is shown in Fig. S6 (see Supplementary Information). Now it is easy to check that, along the $k_c$ axis, the $|\pm 1/2>$ band is two-fold degenerate. Meanwhile, the $|\pm 3/2>$ bands split into two bands which cross with $|\pm 1/2>$ band, forming two TDNPs (Fig. 4b).

Thanks to the unique property of the continuous solid solution in pyrochlore $Tl_2Nb_2O_{6+x}$ ($0 \leq x \leq 1.0$) under oxidation, the TDNPs exist at a high-symmetry line and can move along the line via tuning the oxidation level. The space group of $Tl_2Nb_2O_{6+x}$ is symmorphic ($0 < x \leqslant 0.5$) and the TDNPs here are protected by the rotation symmetry. It is different from the TDNP emerging at high-symmetry point, which are protected by nonsymmorphic symmetries.[24]

Finally, we study the SOC band structure and evolution of the Wannier charge centers of $Tl_2Nb_2O_6$ (*R3m*, No. 160) to make a comparison with those of $Tl_2Nb_2O_{6.5}$ (see Supplementary Information Fig. S7, S8). Owing to the lacking of extra O', all Tl in $Tl_2Nb_2O_6$ are in +1 valence state. The $Tl_2Nb_2O_6$ system shows trivial insulating state with an indirect narrow band gap.

It is noted that $Tl_2Ta_2O_{6+x}$[35] has the same chemical and physical properties as $Tl_2Nb_2O_{6+x}$, as well as the band topology. $La_2Hf_2O_7$ is found to be a QCP semimetal in GGA calculation and it becomes a topological crystalline insulator in GGA+SOC calculation, which are the same as those in topological material database Materiae, and other similar databases.[41-43] The effective SOC splitting in $La_2Hf_2O_7$ is found to be opposite to that of $Tl_2Nb_2O_7$, while the absence of valence variation in La ions makes the oxygen content hard to be tuned in $La_2Hf_2O_7$.

**CONCLUSION**

In summary, we propose that a pyrochlore oxide $Tl_2Nb_2O_{6+x}$ with continuous oxidation level x can host various topological phases, which is realized by a change of valence state of Tl from 1+

to 3+ and the displacement of its atomic position. $Tl_2Nb_2O_7$ with x = 1 is a semimetal with QCP due to cubic symmetry. When a small in-plane tensile strain is applied, a nodal line appears in the non-SOC case, which is protected by the inversion and time-reversal symmetries. When SOC is taken into account, $Tl_2Nb_2O_7$ could harbour bulk Dirac points with gapless topological surface states under a small compressive in-plane strain, or a TI under the expansion case. For $Tl_2Nb_2O_{6.5}$ with x = 0.5 where inversion symmetry is absent, a couple of intrinsic triply degenerate nodal points exist and are protected by time-reversal and $C_{3v}$ symmetries. For $Tl_2Nb_2O_6$ with x = 0, it is a narrow gap semiconductor with trivial topology. On the experimental aspect, the successful fabrication of crystals of $Tl_2Nb_2O_{6+x}$ series[35,36] makes it feasible to observe these topological features as we proposed. Our work can widen the knowledge of TEMs in oxide materials. Furthermore, the realization of different topological states in one series can stimulate the study on the coupling among them, which may generate new physics or interesting transport properties, as well as interacting topological phases.

## COMPUTATIONAL METHODS

We have performed first-principles calculations within density functional theory (DFT), using the Vienna *ab initio* simulation package (VASP).[48,49] Exchange-correlation potential is treated within the generalized gradient approximation (GGA) in the form of Perdew-Burke-Ernzerhof (PBE).[50] BZ is sampled with k-point meshes of $12 \times 12 \times 12$ for self-consistent electronic structure calculations. The spin-orbit coupling is included self-consistently. The simulation of uniaxial strain along [001] is simulated by fixing the experimental volume with the ratio $a/c$ tuned. Here, $a$ is lattice constant along $x/y$ direction, while $c$ is the lattice constant along $z$ direction ([001] direction). The nonlocal Heyd-Scuseria-Ernzerhof (HSE06) hybrid functional calculation is carried out to remedy the possible underestimation of band gap and overestimation of band inversion.[51,52] To calculate $Z_2$ invariant, surface states and nodal line states of the system, the maximally localized Wannier functions (MLWF)[53,54] are introduced into WannierTools[55] and a tight-binding model was constructed.

**ACKNOWLEDGEMENTS**

This work was supported by the National Key Research and Development Program of China (Nos. 2017YFA0304700, 2017YFA0303402, 2016YFA0300600 and 2018YFA0305700), the National Natural Science Foundation of China (Nos. 11504051, 11674077, 11604273, 11704117, and 11674369), and Natural Science Foundation of Fujian Province of China (No. 2018J06001). C. F. and H. W. are also supported by the Science Challenge Project (No. TZ2016004), Beijing Natural Science Foundation (Z180008) and the K. C. Wong Education Foundation (Grant No. GJTD-2018-01).

## AUTHOR CONTRIBUTIONS

W.Z. carried out the first-principles calculations; K.L. and R.Y. constructed the $k \cdot p$ effective model. Z.C. and Z.Z. contributed to analyzing the data. W.Z., R.Y. and H.W. wrote the paper. H.W. supervised the project. All authors participate in research discussion.

## ADDITIONAL INFORMATION

Supplementary information accompanies the paper on the *npj Computational Materials* website.

## COMPETING FINANCIAL INTERESTS

The authors declare no competing financial interests.

## FIGURES AND FIGURE LEGENDS

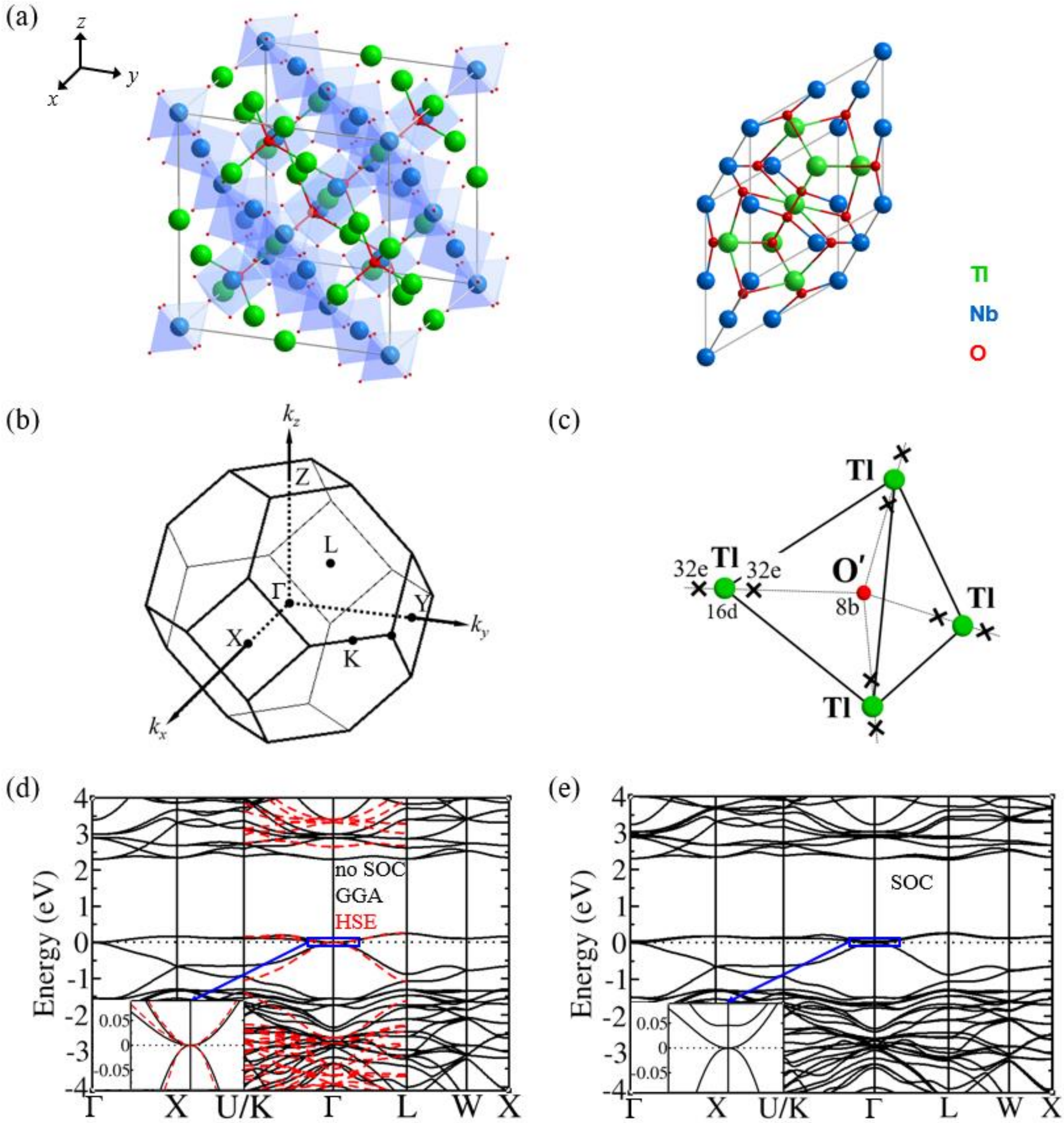

**Fig. 1** $Tl_2Nb_2O_7$ system. **a** Conventional unit cell constructed by a three-dimensional corner-sharing network of $NbO_6$ octahedral (left panel). The related primitive unit cell (right panel). **b** Schematic of the bulk first Brillouin zone (BZ). **c** Tetrahedron constructed by four Tl atoms surrounding O' atom. The tetrahedron corners represent position 16*d*, while crosses represent position 32*e*. **d** Non-SOC band structures calculated within GGA (black solid lines) and hybrid functional HSE06 (red dotted lines). **e** SOC band structure calculated within GGA.

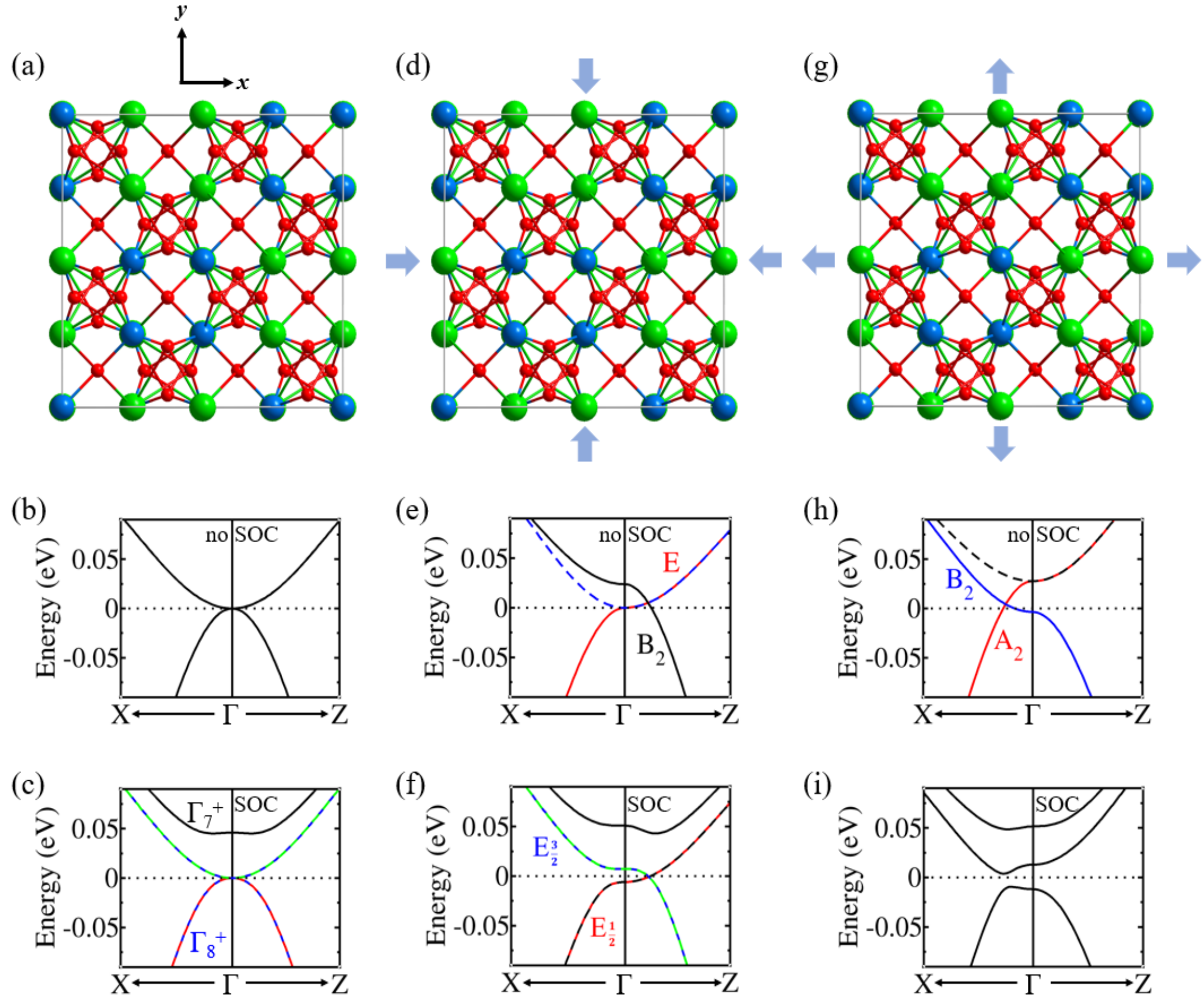

**Fig. 2** Strained $Tl_2Nb_2O_7$ system. **a** Top view of the crystal structure, **b** non-SOC and **c** SOC energy bands for the structure without strain. **d** Top view of the crystal structure, **e** non-SOC and **f** SOC energy bands for the structure with strain of -1% (inplane compression). **g** Top view of the crystal structure, **h** non-SOC and **i** SOC energy bands for the structure with strain of 1% (inplane expansion).

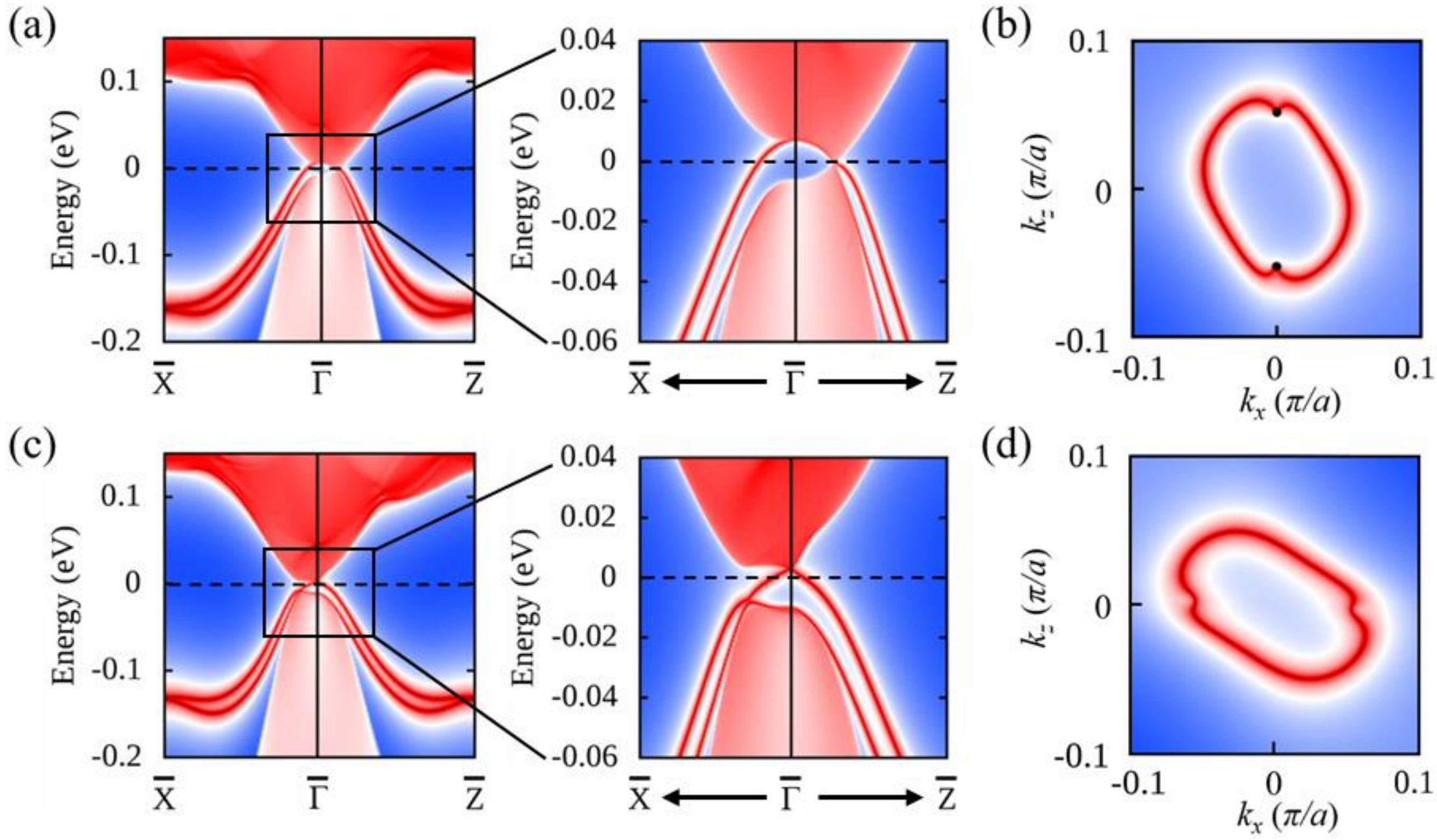

**Fig. 3** Topological properties of strained $Tl_2Nb_2O_7$ with SOC on (010) surface. Structure with inplane strain of -1% (compression): **a** 3D Dirac point and topological surface states both appear. Bulk states and surface states near Fermi level are enlarged and shown on the right panel. **b** Fermi surface of surface states with Fermi energy at -0.737 meV where Dirac points locate. Two project Dirac nodes are represented as black dots. Structure with inplane strain of 1% (expansion): **c** Calculated topological surface states. Bulk states and surface states near Fermi level are enlarged and shown on the right panel. **d** Fermi surface of surface states with Fermi energy at -6 meV.

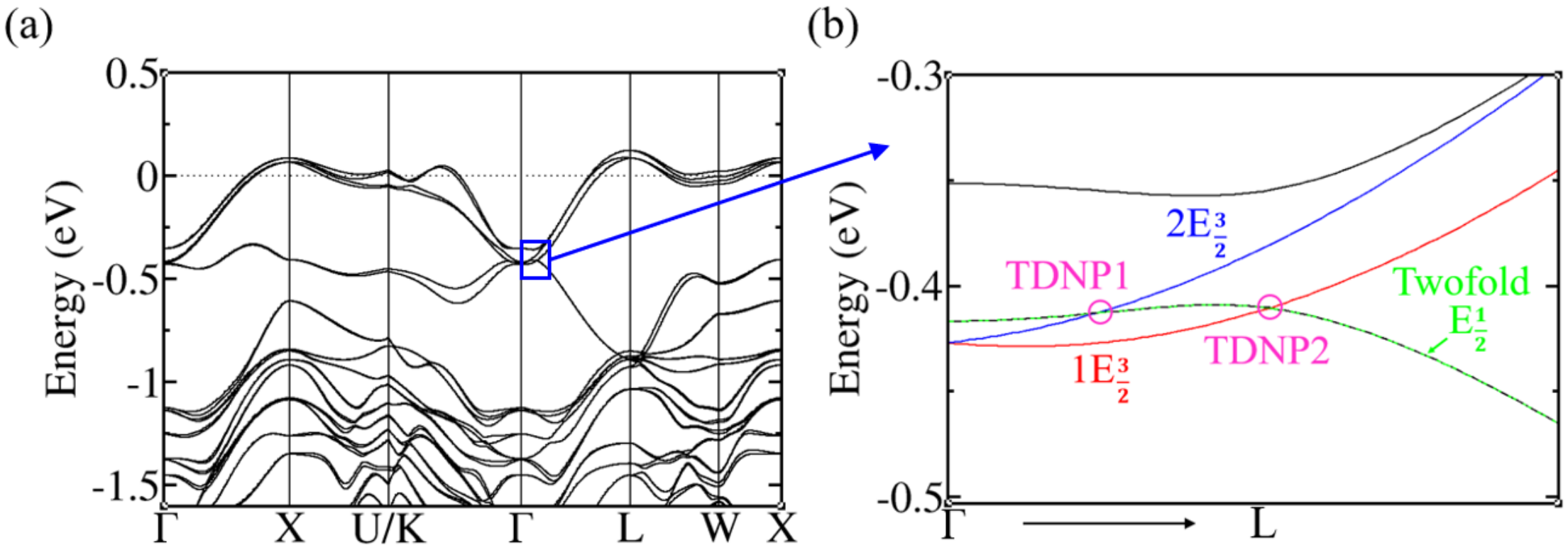

**Fig. 4** Band structures of $Tl_2Nb_2O_{6.5}$. **a** SOC band structure calculated within GGA. **b** Enlarge band structure along Γ-L around TDNP in **a**.